\documentclass[12pt]{iopart}

\ifx\pdfoutput\undefined
   \usepackage[dvips]{graphicx}
   \else
   \usepackage[pdftex]{graphicx}
   \fi
   \usepackage{epstopdf}

\usepackage{graphics}
\usepackage{graphicx}
\usepackage{epsfig}

\usepackage[square,comma,numbers,sort&compress]{natbib}

\begin{document}

\DeclareGraphicsExtensions{.pdf,.gif,.jpg}

\title[LISA Sensitivity Curves]{The construction and use of LISA sensitivity curves}

\author{Travis Robson\dag, Neil J. Cornish\dag$\;$ \& Chang Liu\ddag$\sharp$\dag}

\address{\dag\ eXtreme Gravity Institute, Department of Physics, Montana State University, Bozeman,
MT 59717, USA}

\address{\ddag\ CAS Key Laboratory of Theoretical Physics, Institute of Theoretical Physics, Chinese Academy of Sciences, Beijing 100190, China}
\address{$\sharp$ School of Physical Sciences, University of Chinese Academy of Sciences, Beijing 100049, China}

\bibliographystyle{iopart-num}

\begin{abstract} 
The Laser Interferometer Space Antenna (LISA) will open the mHz band of the gravitational wave spectrum for exploration. Sensitivity curves are a useful tool for surveying the types of sources that can be detected by the LISA mission.
Here we describe how the sensitivity curve is constructed, and how it can be used to compute the signal-to-noise ratio for a wide range of binary systems. We adopt the 2018 LISA Phase-0 reference design parameters. We consider both sky-averaged sensitivities, and the sensitivity to sources at particular sky locations. The calculations are included in a publicly available {\em Python} notebook.
\end{abstract}

We describe the construction and use of LISA sensitivity curves, the computation of signal-to-noise ratios, and how to plot signal strengths against the sensitivity curve. Figure 1 shows an example of a sensitivity/source plot taken from the LISA L3 mission proposal~\cite{Audley:2017drz}. The idea, in plotting signal and noise curves in this manner, is that the height a signal is above the sensitivity curve indicates how loud it will be.

\begin{figure}[h]
\includegraphics[scale=0.8]{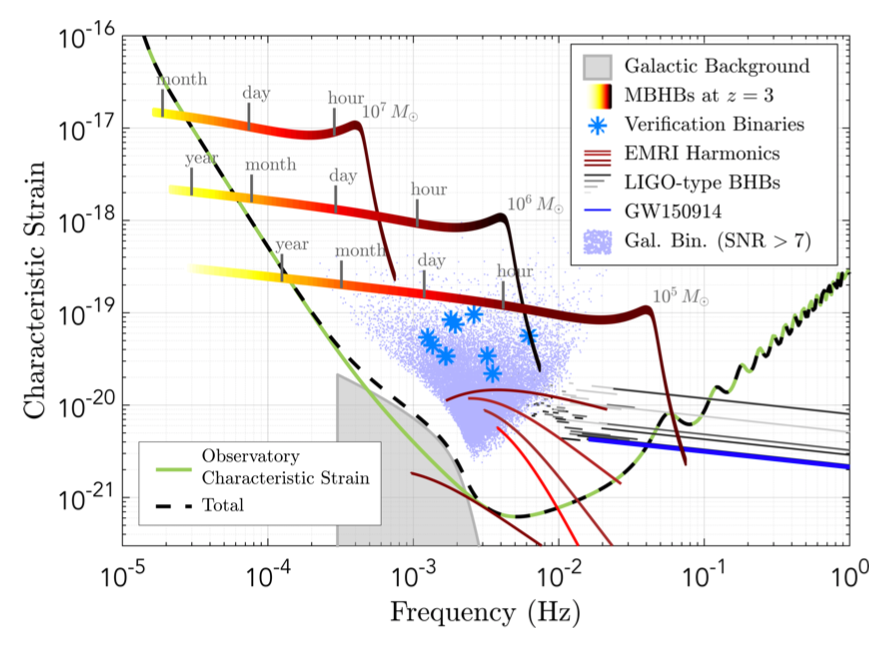} 
\centering
\caption{A plot taken from the LISA L3 mission proposal showing the expected sensitivity (green line) and a variety of possible sources (various colors) in units of dimensionless characteristic strain.}
\end{figure}

The literature on this topic can be very confusing, with a profusion of conventions and notation. Unfortunately, some of the choices that have now become standard are misleading, but it is probably too late to change the conventions now. Sensitivity curves are useful for making a quick assessment of what signals may be detectable. While not used for actual data analysis, the sensitivity curve, and signal representations that are shown with them, are designed to represent the quantities that are used in the data analysis. See Ref.~\cite{Moore:2014lga} for a review of gravitational wave sensitivity curves. While useful, the sky-averaged sensitivity can be misleading as there is often significant variation in the sensitivity with sky location. To this end, we also provide expressions and tools for computing the signal-to-noise ratio as a function of sky location, averaged over inclination and polarization angles.

Python code and a Jupyter notebook for generating the results shown in this document can be downloaded from GitHub~\cite{github}.
 The Jupyter notebook can be edited and executed directly in your browser without the need to install any software using the MyBinder version~\cite{mybinder}.

\section{Sensitivity Curves}

The LISA sensitivity curve can be well approximated by the equation
\begin{equation}\label{Sn}
S_n(f) = \frac{10}{3 L^2} \left( P_{\rm OMS}(f)  +  \frac{4P_{\rm acc}(f)}{(2\pi f)^4} \right) \left(1 + \frac{6}{10}\left(\frac{f}{f_*}\right)^2 \right) + S_c(f) \, ,
\end{equation}
where $L= 2.5$ Gm, $f_*= 19.09$ mHz, and expressions for $P_{\rm OMS}(f)$, $P_{\rm acc}(f)$ and $S_c(f)$ are given in equations (\ref{oms}), (\ref{acc}) and (\ref{gf}) below. Here we explain how this curve is computed and how it can be used (and sometimes mis-used).

The simplest type of sensitivity curve, and the one used by the ground-based detector community, is the power spectral density of the detector noise $P_n(f)$, or the amplitude spectral density 
$\sqrt{P_n(f)}$. The mean-squared noise in the frequency band $[f_1,f_2]$ is just the integral of $P_n(f)$ over that band. But for a detector like LISA, where signals may have wavelengths that are shorter than the arms of the detector, it is conventional to include the ensuing arm-length penalty in the sensitivity curve~\cite{Larson:1999we}.  The strain spectral sensitivity is then defined in terms of the square root of the effective noise power spectral density
\begin{equation}
S_n(f) = \frac{P_n(f)}{{\cal R}(f)},
\end{equation}
where ${\cal R}(f)$ is the sky and polarization averaged signal response function of the instrument. The signal response function ${\cal R}(f)$ relates the power spectral density of the incident gravitational wave signals to the power spectral density of the signal recorded in the detector. As such, it might have been more logical to include this factor in the expression of the signals, but early on it was decided to apply the inverse of this factor to the noise power to define a sensitivity curve - c'est la vie.

The response function can be computed by working in the frequency domain, where the gravitational wave amplitude in the detector, $\tilde{h}(f)$, is related to the plus and cross gravitational wave amplitudes via
\begin{equation}
\tilde{h}(f) = F^+(f) \tilde{h}_+(f) + F^\times(f) \tilde{h}_\times(f) \, ,
\end{equation}
where $F^+(\theta,\phi,\psi,f)$ and $F^\times(\theta, \phi,\psi,f)$ are the (complex) frequency dependent detector response functions, which depend on the sky location $(\theta,\phi)$ and polarization angle $\psi$ of the source.  The sky/polarization averaged spectral power of the signal in the detector, $\langle \tilde{h}(f) \tilde{h}^*(f)\rangle$ is related to the raw spectral signal power $|\tilde{h}_+(f)|^2 +| \tilde{h}_\times(f)|^2$ by the response function:
\begin{eqnarray}\label{Rdef}
&& \langle \tilde{h}(f) \tilde{h}^*(f)\rangle = \langle F^+(f) F^{+*}(f)\rangle |\tilde{h}_+(f)|^2 +  \langle F^\times(f)F^{\times*}(f)  \rangle |\tilde{h}_\times(f)|^2  \nonumber \\
&& \quad = {\cal R}(f)\left(|\tilde{h}_+(f)|^2 +| \tilde{h}_\times(f)|^2\right)
\end{eqnarray}
where ${\cal R}(f) =  \langle F^+(f) F^{+*}(f) \rangle =  \langle F^\times(f) F^{\times*}(f) \rangle$, and the angle brackets indicate the sky/polarization average
\begin{equation}
\langle  X \rangle  \equiv \frac{1}{4\pi^2} \int_0^\pi d\psi \int_0^{2 \pi} d\phi \int_0^\pi  X \, \sin\theta \, d\theta.
\end{equation}
For a right-angle interferometer operating in the long wavelength limit, such as LIGO/Virgo, the antenna patterns are real and independent of frequency, and are given by
\begin{eqnarray}
&& F^+ = \frac{1}{2}(1+\cos^2\theta) \cos(2\phi) \cos(2\psi) -\cos\theta \sin2\phi \sin2\psi \nonumber \\
&& F^\times = \frac{1}{2}(1+\cos^2\theta) \cos(2\phi) \sin(2\psi) + \cos\theta \sin2\phi \cos2\psi \,\,.
\end{eqnarray}
For LIGO we have
\begin{equation}
{\cal R} =  \langle {F^+}^2\rangle = \langle {F^\times}^2\rangle = \frac{1}{32} \int_{-1}^{1} (1+6 x^2 + x^4) dx = \frac{1}{5} \, .
\end{equation}
In the LIGO literature this factor is applied to the signals, leaving the sensitivity curve to be just the power spectral density of the noise. The full expressions for $F^+(f)$ and $F^\times(f)$ for the Michelson-style interferometry signals for LISA are much more complicated than those for LIGO (they are given in equations (5), (6), (16) and (17) of Ref.~\cite{Cornish:2001bb}.) For a 3-arm LISA, there are two independent channels for $f < f_*$ and three for $f > f_*$, where $f_* = c/(2\pi L)$ is the transfer frequency~\cite{Prince:2002hp}. For the current LISA design, $L= 2.5$ Gm, and $f_*= 19.09$ mHz. The standard convention is to define ${\cal R}(f)$ as being summed over the plus and cross channels. For sources that have frequency components $f > f_*$, it is more accurate to consider the 3-channel expressions given in Ref.~\cite{Prince:2002hp}. The full expression for ${\cal R}(f)$ is not known in closed form, but to leading order is given by
\begin{equation}
{\cal R}(f) = \frac{3}{10} -\frac{507}{5040}\left(\frac{f}{f_*}\right) + \dots
\end{equation}
The first term, $3/10$, is a factor of $2 \times \sin^2(60^\circ)= 3/2$ larger than the corresponding LIGO result due to the LISA having two low-frequency channels, and arms that make an angle of $60^\circ$, as opposed to the $90^\circ$ angle for LIGO. The full expression for ${\cal R}(f)$ has to be computed numerically~\cite{Larson:1999we}, and has the form shown in Figure 2. The transfer function can be well-fit by the curve
\begin{equation}\label{Rfit}
{\cal R}(f) = \frac{3}{10}  \frac{1} { \left(1 + 0.6 (f/f_*)^2 \right)} \,\,.
\label{eq:approxR}
\end{equation}
Note that many publications quote the number $3/20$ for the low frequency limit of $R(f)$. The factor of two larger value quoted in eqn.~(\ref{eq:approxR}) comes from summing over the two independent low-frequency data channels.

\begin{figure}[ht]
\includegraphics[scale=0.8]{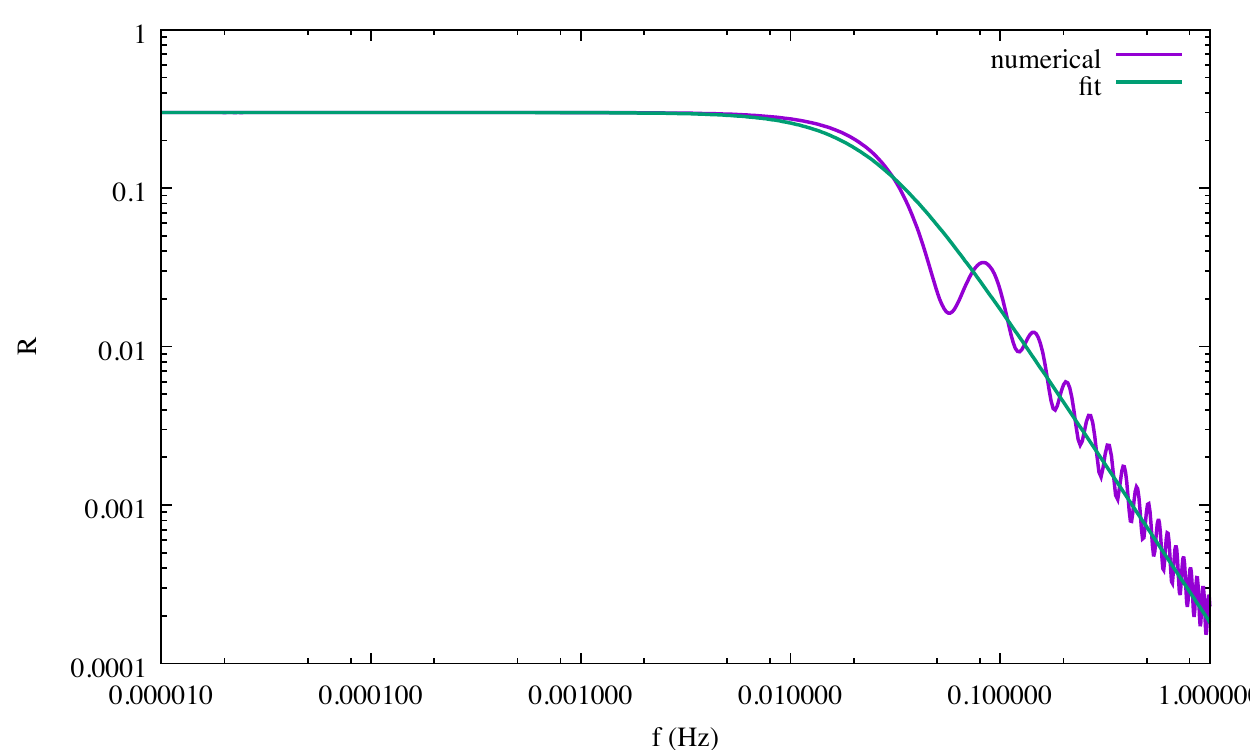} 
\centering
\caption{The signal transfer function ${\cal R}(f)$ for the combination of two Michelson-style LISA data channels, and the analytic fit from equation (\ref{Rfit}).}
\end{figure}

The current ``official'' model for the power spectral density of the LISA noise $P_n(f)$ is based on the Payload Description Document, and is referenced in the ``LISA Strain Curves''  document LISA-LCST-SGS-TN-001.  The single-link optical metrology noise is quoted as
\begin{equation}\label{oms}
P_{\rm OMS} = (1.5 \times 10^{-11} \, {\rm m})^2 \left( 1+ \left(\frac{ 2\, {\rm mHz}}{f}\right)^4 \right)\, {\rm Hz}^{-1}\, ,
\end{equation}
and the single test mass acceleration noise is quoted as
\begin{equation}\label{acc}
P_{\rm acc} = (3 \times 10^{-15} \, {\rm m}\, {\rm s}^{-2})^2 \left( 1+ \left(\frac{ 0.4\, {\rm mHz}}{f}\right)^2 \right) \left( 1+ \left(\frac{f}{ 8\, {\rm mHz}}\right)^4 \right) \, {\rm Hz}^{-1}\, .
\end{equation}
The total noise in a Michelson-style LISA data channel is then~\cite{Cornish:2001bb}
\begin{equation}\label{Pn}
P_n(f) = \frac{P_{\rm OMS}}{L^2}  + 2(1+\cos^2(f/f_*)) \frac{P_{\rm acc}}{(2\pi f)^4 L^2} \, .
\end{equation}
Note that the Michelson-style response has four contributions from the optical metrology noise and sixteen from the test mass acceleration noise. We convert from displacement to strain by dividing by the round-trip light travel distance $2L$, so, for example, the factor of $4 P_{\rm OMS}$ gets divided by $(2L)^2$, leading to the expression seen in (\ref{Pn}). The same factor of $1/(2L)$ is also applied to the path-length change caused by the gravitational wave, so it cancels out in the likelihood function and the SNR, and the choice to divide by $2L$ is an unimportant convention.
A good analytic model for the sensitivity curve that is sufficient for most purposes is given by combining (\ref{Rfit}) and (\ref{Pn}):
\begin{equation}\label{Sn2}
S_n(f) = \frac{10}{3 L^2} \left( P_{\rm OMS}  + 2(1+\cos^2(f/f_*)) \frac{P_{\rm acc}}{(2\pi f)^4} \right) \left(1 + \frac{6}{10}\left(\frac{f}{f_*}\right)^2 \right) \, .
\end{equation}

\begin{figure}[th]
\includegraphics[scale=0.8]{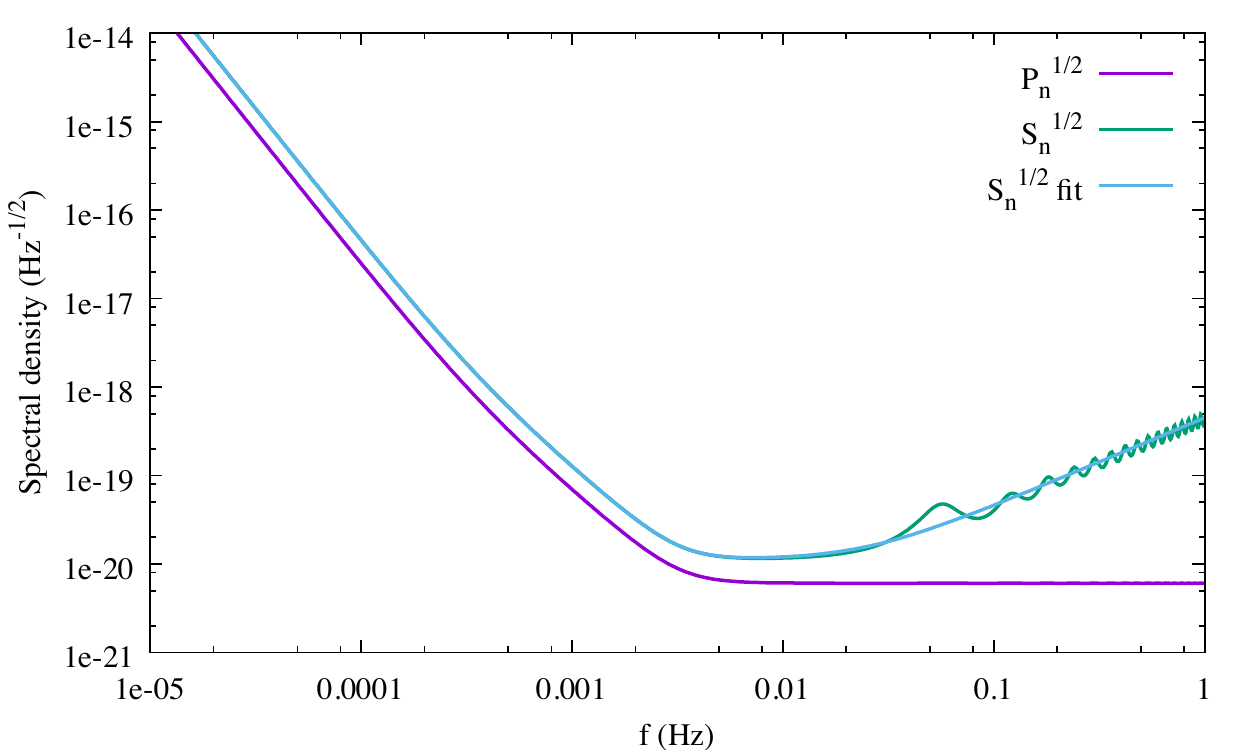} 
\centering
\caption{The amplitude spectral density of the noise, and the corresponding sensitivity curve, found by dividing $P_n(f)$ by ${\cal R}(f)$. The analytic fit to $S_{n}(f)$ given in equation (\ref{Sn}) is also shown.}
\end{figure}

In addition to the instrument noise, unresolved galactic binaries will act as an effective noise source (though one that is not stationary). The galactic confusion noise goes down as the mission progresses and more foreground sources are removed. Estimates for the confusion noise using the new LISA design are given in Ref.~\cite{Cornish:2017vip}, and are well fit by the function
\begin{equation}\label{gf}
S_c(f) = A \,  f^{- 7/3}\, e^{-f^{\alpha} + \beta  f  \sin(\kappa f) } \left[1+{\rm tanh}(\gamma(f_k-f))\right]  \, {\rm Hz}^{-1}
\end{equation}
with fit parameters given in Table 1. Note that the amplitude quoted here is half the value quoted in Ref.~\cite{Cornish:2017vip} since here we are using two-channel sensitivity curves. The full sensitivity curve is found by adding $S_c(f)$ to $S_n(f)$.
\begin{table}
\begin{center}
	\begin{tabular}{ | c | c | c | c | c | }
		\hline
		& 6 mo & 1 yr & 2 yr & 4 yr \\ [0.5ex]
		\hline \hline
		$\alpha$&0.133&0.171&0.165&0.138\\
		$\beta$&243&292&299&-221 \\
		$\kappa$&482&1020&611&521\\
		$\gamma$&917&1680&1340&1680\\
		$f_{k}$&0.00258&0.00215&0.00173&0.00113\\
		\hline\hline
	\end{tabular}
\end{center}
\caption{Parameters of the analytic fit the Galactic confusion noise as described by equation (\ref{gf}). The amplitude $A$ has been fixed to $9\times 10^{-45}$. Note that the knee frequency $f_{k}$ decreases with observation time and $\gamma$ increase with observation time, leading to a steeper drop off in confusion noise.}
\end{table}

\begin{figure}[th]
\includegraphics[scale=0.8]{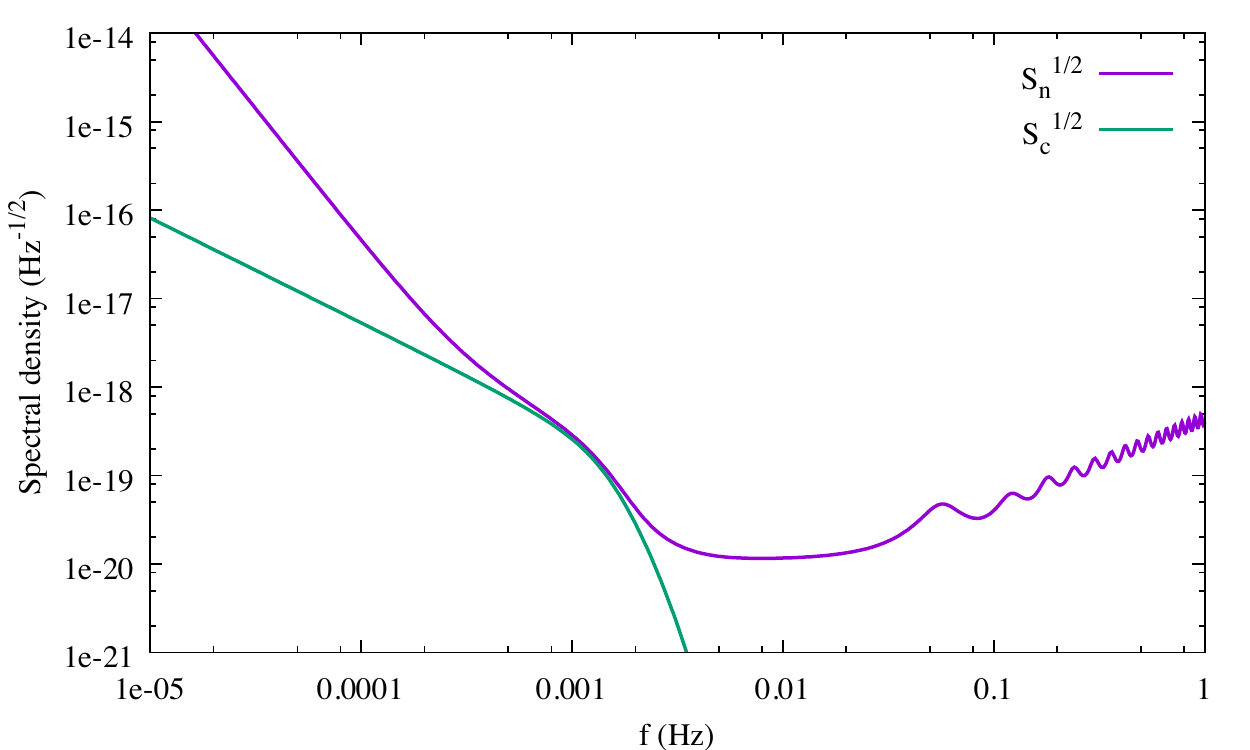} 
\centering
\caption{The amplitude spectral density of the galactic noise, $S_c^{1/2}$, and the full sensitivity curve combining the instrument noise and the galactic confusion noise, $S_n^{1/2}$, for a 4-year mission lifetime.}
\end{figure}

Figure 4 shows the contribution of the galactic confusion noise assuming a 4-year mission, along with the updated sensitivity curve that includes the confusion noise. Note that the confusion noise shown here is an average value - in practice it will vary over a year as the LISA antenna pattern sweeps across the galaxy.

\section{Binary Sources}

The majority of LISA sources will be binaries of various masses and mass ratios. For simplicity we will focus here on quasi-circular, non-spinning comparable mass binaries, and only consider the dominant quadrupole harmonic. Extreme mass ratio binaries, which may be highly eccentric, require a more involved treatment. While we ignore spin, the signal-to-noise ratios we compute should be good to within a factor of two or so for spinning systems. Our model for the waveforms is then
\begin{eqnarray}
&& \tilde{h}_+(f)=  A(f) \frac{(1+\cos^2\iota)}{2} \, e^{i\Psi(f)}\nonumber \\
&& \tilde{h}_\times(f) = iA(f)   \cos\iota \, e^{i\Psi(f)} \, ,
\end{eqnarray}
where $\iota$ describes the inclination of the orbit relative to the line of sight, and $A(f)$ and $\Psi(f)$ are the amplitude and phase of the wave. To compute the sky/polarization averaged SNR we only need to know $A(f)$, and in some cases, how the frequency evolves with time, $f(t)$. 

Earlier we related the sky and polarization averaged power spectral density of the signal to the power spectral density seen in the detector via equation (\ref{Rdef}). For binary systems it is natural to extend the angle averaging to include the inclination angle:
\begin{equation}
\langle \tilde{h}(f) \tilde{h}^*(f)\rangle =  {\cal R}(f) A^2(f) \frac{1}{2} \int_{-1}^{1} \left( \frac{\left(1+x^2\right)^2}{4} + x^2\right) dx = \frac{4}{5} {\cal R}(f) A^2(f) \, .
\end{equation}
Note that for LIGO we recover the well-known pre-factor $\sqrt{(4/5) {\cal R}(f)}= 2/5$ that is applied to the GW amplitude to account for averaging over the source location and orientation~\cite{Finn:1992xs}.

The amplitude signal-to-noise ratio $\rho$ for a  deterministic signal $\tilde{h}(f)$ is given by
\begin{equation}
\rho^2 = 4 \int 
\frac{\vert \tilde{h}(f) \vert^2}{P_n(f)} \, df 
= 4  \int_{f=0}^\infty 
 \frac{f  \vert  \tilde{h}(f) \vert^2}{P_n(f)} \, d(\ln f) \, .
\end{equation}
Averaging over sky location, inclination and polarization we have
\begin{equation}
\overline{\rho^2} 
=  \frac{16}{5}  \int 
\frac{ f A^2(f)} {P_n(f)} \,d(\ln f) = \frac{16}{5}  \int 
\frac{  (2 f  T) S_h(f)}{S_n(f)} \, d(\ln f) \, , 
\end{equation}
Where $T$ is the observation time and $S_h(f)$ is the one-sided, angle averaged, power spectral density of the signal, 
\begin{equation}
S_h(f) = \frac{A^{2}(f)}{2 T} \, .
\end{equation}
If you took a Fourier transform of the data, $d = h +n$, then ignoring any correlations between the signal and the noise, the power spectral density of the data would equal to $S_d(f) = S_h(f) + P_n(f)$. In other words, $S_h(f)$ is the power spectral density of the signal. The factor of $(2 f  T)$ that appears in the expression for the optimal signal-to-noise shows that the signal is effectively boosted relative to the noise by using templates to coherently extract the signal. Rather than plotting the signal power directly (which often lies below the sensitivity curve), the convention is to plot  $h_{\rm eff}^2=16 f (2 f  T) S_h(f)/5$, to account for the boost we get from the coherent signal extraction. 

For the waveform model we use the original phenomenological inspiral-merger-ringdown (IMR) model, known as PhenomA~\cite{Ajith:2007qp}. While more accurate models now exist, such as the latest PhenomP model~\cite{Hannam:2013pra,Schmidt:2014iyl}, which includes spin-precession, PhenomA is good enough for making graphs and estimating SNRs. The PhenomA amplitude is given by
\begin{eqnarray}\label{PhA}
A(f) &\equiv& \sqrt{\frac{5}{24}}\frac{( G{\cal  M}/c^3)^{5/6}f_0^{-7/6}}{\pi^{2/3}(D_L/c)}\,
\left\{ \begin{array}{ll}
\left(\frac{f}{f_0}\right)^{-7/6} 
& \textrm{if~~$f <  f_0$} \\ \\
\left(\frac{f}{f_0}\right)^{-2/3} & \textrm{if~~$f_0 \leq f < f_1$}  \\  \\
w \, {\cal L}\left(f,\,f_1,\,f_2\right) &\textrm{if~~$f_1 \leq  f <  f_3$}\,,  \\
\end{array} \right.\\  \nonumber 
\end{eqnarray}
where
\begin{equation}
f_k \equiv \frac{a_k \eta^2 + b_k \eta + c_k}{\pi (G M/c^3)}\, ,
\end{equation}
\begin{equation}
{\cal L}(f,f_1,f_2) \equiv \left(\frac{1}{2 \pi}\right)
\frac{f_2}{(f-f_1)^2+f_2^2/4} \,,
\label{eq:Lorenzian}
\end{equation}
and
\begin{equation}
w \equiv \frac{\pi f_2}{2} \left(\frac{f_0}{f_1}\right)^{2/3}\,.
\end{equation}
Here $M=m_1+m_2$ is the total mass, $\eta = m_1 m_2/M^2$ is the symmetric mass ratio and ${\cal M}= (m_1 m_2)^{3/5}/M^{1/5}$ is the chirp mass. The $G$'s and $c$'s have been included for those that are not used to working in natural units. Note that the combinations $G M/c^3$ and $D_L/c$ both have units of time. A useful number to remember is that the mass of the Sun, $G M_\odot /c^3$, is approximately 5 microseconds in natural units.
The coefficients for the transition frequencies $f_k$ are given in Table 2. Roughly speaking, $f_0$ is the merger frequency, $f_1$ is the ringdown frequency, $f_2$ is decay-width of the ringdown and $f_3$ is the cut-off frequency.

\begin{table}[tbh]
    \begin{center}
        \begin{tabular}{cccccccc}
            \hline
            \hline
             &\vline& \multicolumn{1}{c}{$a_k$} &\vline& \multicolumn{1}{c}{$b_k$} &\vline& \multicolumn{1}{c}{$c_k$} \\
            \hline
            $f_0$ &\vline& 2.9740$\times 10^{-1}$   &\vline& 4.4810$\times 10^{-2}$    &\vline&  9.5560$\times 10^{-2}$    \\
            $f_1$ &\vline& 5.9411$\times 10^{-1}$   &\vline& 8.9794$\times 10^{-2}$    &\vline& 1.9111$\times 10^{-1}$     \\
            $f_2$ &\vline& 5.0801$\times 10^{-1}$   &\vline& 7.7515$\times 10^{-2}$    &\vline&  2.2369$\times 10^{-2}$    \\
            $f_3$  &\vline& 8.4845$\times 10^{-1}$   &\vline& 1.2848$\times 10^{-1}$    &\vline&  2.7299$\times 10^{-1}$    \\
            \hline
            \hline
        \end{tabular}
        \caption{Polynomial coefficients of the transition frequencies.}
        \label{tab:polCoeffsAmpParams}
    \end{center}
\end{table}

\begin{figure}[th]
\includegraphics[scale=1.0]{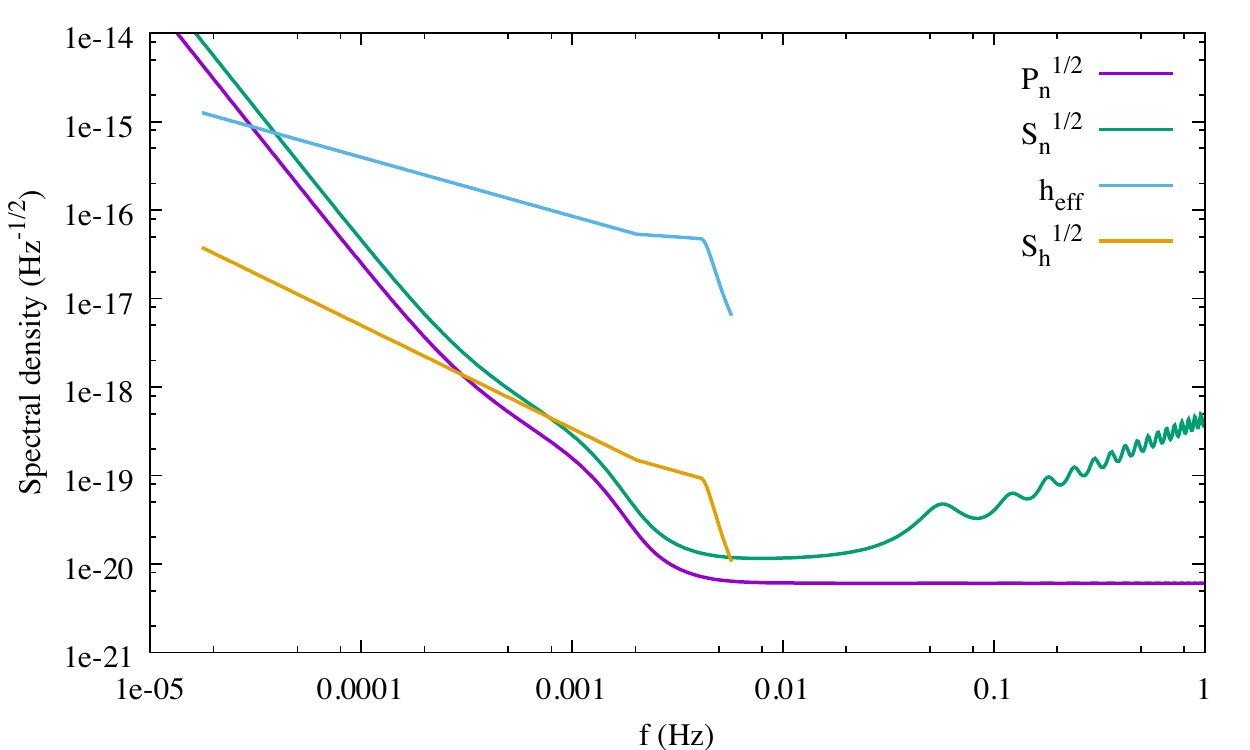} 
\centering
\caption{The amplitude spectral density of the noise $\sqrt{P_n}$, and the amplitude sensitivity curve $\sqrt{S_n}$ are plotted against the raw strain spectral density $\sqrt{S_h}$ and the effective strain spectral density $h_{\rm eff}$ for an equal mass black hole binary at $z=3$ with source frame total mass $M = 10^6 \, M_{\odot}$. This system is so bright that even its raw amplitude will be visible in the detector. However, the  effective amplitude $h_{\rm eff}$ that appears in the numerator of the SNR calculation better communicates the true brightness of the source. The area between the $h_{\rm eff}$ curve and the $S_n^{1/2}$ curve roughly corresponding to the optimal SNR of 2626. Note that this graph differs slightly from the one shown in Figure 1, which plots dimensionless characteristic strain $h_c(f) = \sqrt{f S(f)}$ rather than strain spectral density $\sqrt{S}$.}
\end{figure}

The final ingredient we need for computing the SNR is the frequency range covered by the signal. For comparable mass black holes, with $M > 10^4 M_{\odot}$, the signal will sweep across the LISA band and merge in less than the mission lifetime. However, for lower mass systems, such as stellar origin black holes that will merger in the LIGO band a decade or so later, or for white dwarf binaries, which may be millions of years from merger, we need to specify the start and end frequencies for the SNR integration. To leading post-Newtonian order, the frequency as a function of time is given by
\begin{equation}\label{ft}
f(t) = \frac{1}{ 8 \pi (G{\cal M}/c^3)} \left( \frac{ 5 (G {\cal M}/c^3)}{t-t_c} \right)^{3/8} \, ,
\end{equation}
where $t_c$ is the time of coalescence. For example, an equal mass binary at $z=3$ with a total source-frame mass of $M = 10^6 \, M_{\odot}$ will have a GW frequency of $2.93\times10^{-5}$ Hz one year prior to merger. Note that it is the detector frame mass, $M_z = M(1+z)$, that should be used in equations (\ref{PhA}) and (\ref{ft}). For these high mass systems it makes sense to plot $h_{\rm eff}$ across the entire LISA band, and not worry about setting limits in the SNR integration. Tracks in $\sqrt{S_h}$ and $h_{\rm eff}$ for the aforementioned source are shown in Figure~5. In contrast, a source similar to GW150914 that is 5 years from merger when LISA turns on will sweep from $f=16$ mHz to $f=29$ mHz over the nominal 4 year mission lifetime.

\begin{figure}[th]
\includegraphics[scale=1.0]{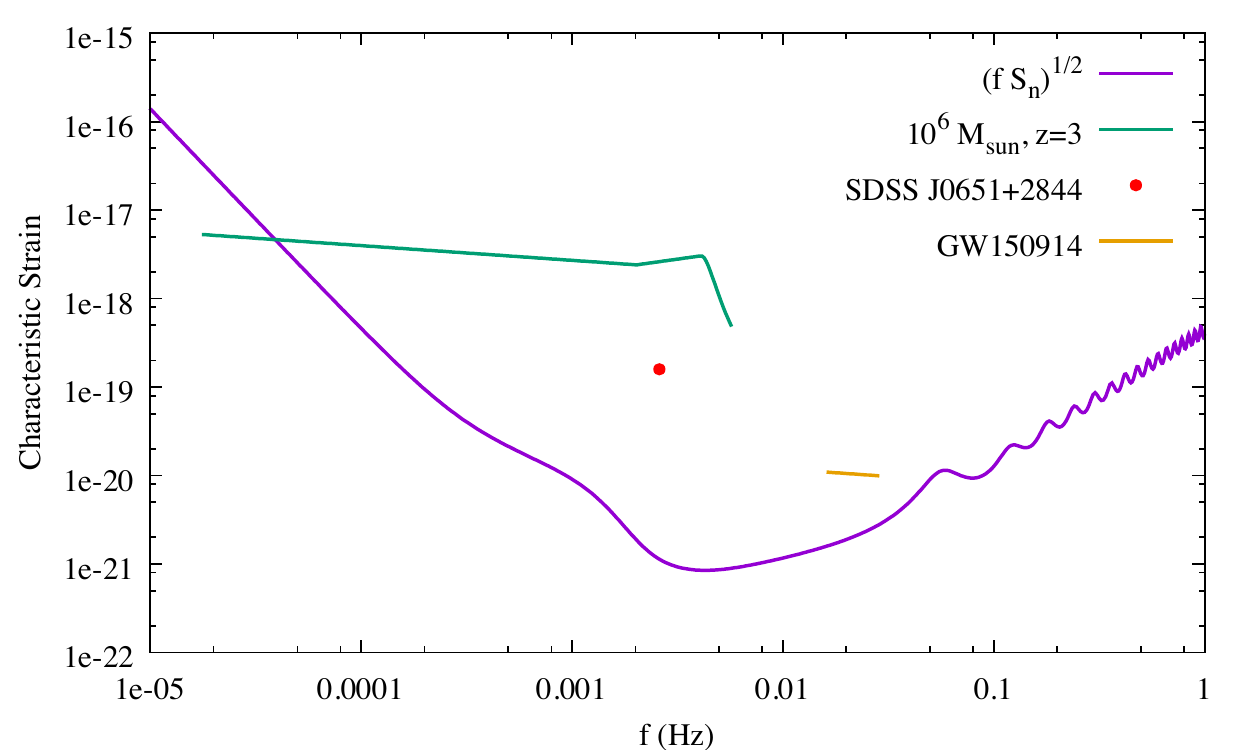} 
\centering
\caption{\label{fig:all3} The sensitivity curve in terms of characteristic strain, $\sqrt{f S_n}$ is compared to three types of signal: an equal mass black hole binary at $z=3$ with source-frame total mass $M = 10^6 \, M_{\odot}$; the galactic verification binary SDSS J0651+2844 observed for 4 years; and a signal similar to the first LIGO detection GW150914 if the LISA observation started 5 years prior to merger and continued for 4 years.}
\end{figure}

For galactic binaries the time to merger is typically very large compared to the mission lifetime, and the frequencies will evolve very little over the course of the mission. Taylor expanding (\ref{ft}) we find
\begin{equation}
f(t) = f_{\rm in} + \frac{96}{5} \pi^{8/3} (G {\cal M}/c^3)^{5/3} f_{\rm in}^{11/3} (t-t_{\rm in}) + \dots \, ,
\end{equation}
where $f_{\rm in}$ is the GW frequency at the start of the observation, at time $t_{\rm in}$. For typical galactic binaries the change in frequency $\Delta f$ during the mission lifetime is so small that it no longer makes sense to the plot the signals as tracks. Rather, the signals are plotted as points with an amplitude $h_{\rm GB}$ that follows from evaluating the SNR integral:
\begin{equation}
\overline{\rho^2} 
=  \frac{16}{5}  \int_{f_{\rm in}}^{f_{\rm in}+\Delta f}
\frac{ A^2(f)} {S_n(f)} \,df  \approx  \frac{16}{5} \frac{\Delta f A^2(f_{\rm in})} {S_n(f_{\rm in})} \equiv  \frac{h^2_{\rm GB}(f_{\rm in})}{S_n(f_{\rm in})}
\end{equation}
where 
\begin{equation}\label{hgb}
h_{\rm GB} = \frac{ 8\,  T^{1/2}  (G {\cal M}/c^3)^{5/3}\pi^{2/3} f_{\rm in}^{2/3}}{5^{1/2} (D_L/c)} \, .
\end{equation}
For example, SDSS J0651+2844 which has $D_L \sim 1$ Kpc, $m_1 \sim  0.5\, M_\odot$, $m_2 \sim 0.25\, M_\odot$, and $f_{\rm in} = 2.6$ mHz, will produce a strain spectral density of 
$h_{\rm GB} = 2.8 \times 10^{18} \, {\rm Hz}^{-1/2}$ and have an SNR of 140 assuming a 4 year mission lifetime. Of course, it not strictly correct to compute angle averaged SNRs for a source with a known sky location and orientation, nor does it make much sense to plot its amplitude against against an all-sky averaged sensitivity curve, but doing so allows us to put all LISA sources on a single graph.

\begin{figure}[th]
\center \includegraphics[scale=1.0]{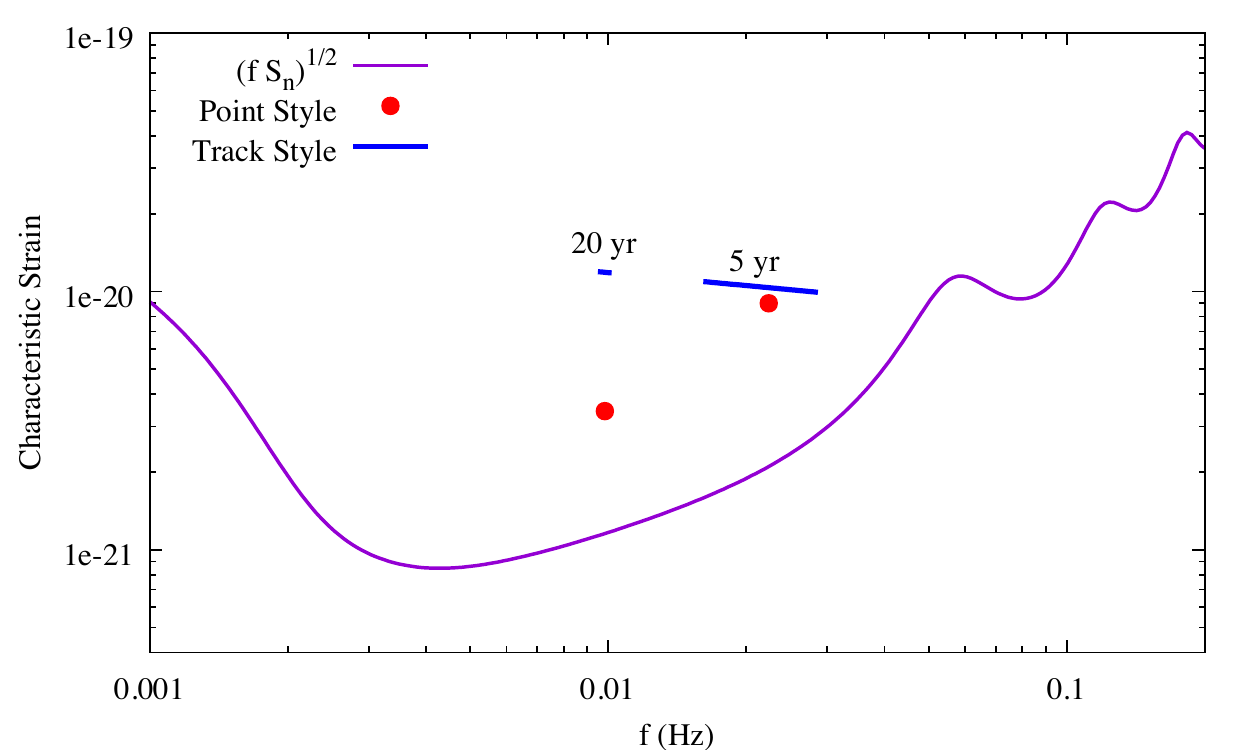} 
\caption{The characteristic strain produced by a GW150914 type system that is either 20 years or 5 years from merger at the beginning of the LISA observation. The track-style representation, where the SNR is estimate from the area under the curve, is compared to the point style representation used for slowly evolving white-dwarf binaries, where the SNR is given by the ratio of the height of the sensitivity curve and the hight of the point.}
\end{figure}

The differing conventions between how slowly evolving and rapidly evolving signals are plotted can be problematic for stellar origin black hole binaries (SOBHBs). For example, if a GW150914 type system was 20 years from merger when LISA started observations, it would be emitting at a gravitational wave frequency of 9.5 mHz, and four years later it would be emitting at 10.4 mHz, producing a track that runs for just $\Delta \ln f = 0.09$. Since the frequency range is so short, the questions becomes do we treat the system as evolving, and plot a track as we do for massive black holes, or do we treat the system as non-evolving, and plot a point as we do for galactic binaries? Figure 7 shows that the two choices paint an inconsistent picture. If the track is longer, such as for a system that is 5 years from merger, the two representations look more consistent.
To arrive at consistent representations, where sources appear at almost the same height when shown as evolving tracks or non-evolving points, we recommend switching from tracks to points when $\Delta \ln f < 0.5$.

\subsection{EMRIs and other complicated signals}

\begin{figure}[th]
\includegraphics[scale=1.0]{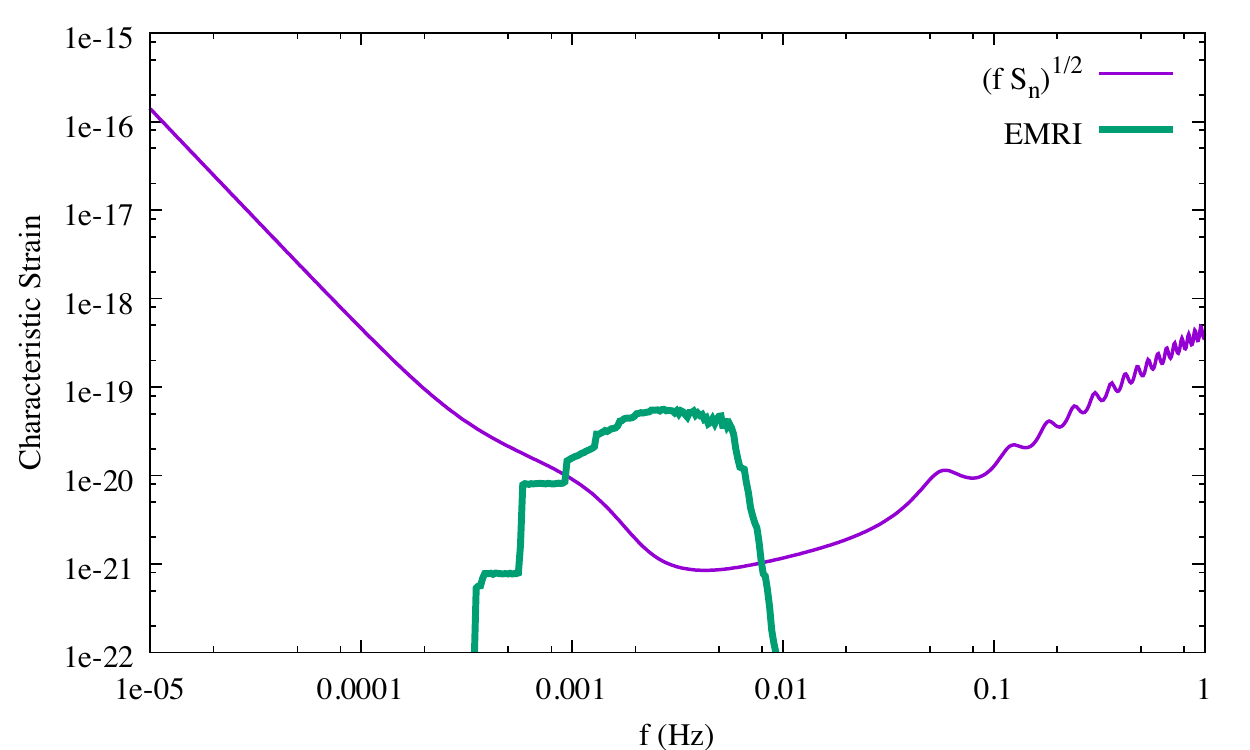} 
\centering
\caption{\label{fig:emri1} The characteristic strain produced by a  $20\, M_\odot$ - $10^6 M_\odot, \chi = 0.5$ EMRI at 4 Gpc.}
\end{figure}

Some sources produce signals that can not be accurately described by simple frequency domain models. Extreme Mass Ratio Inspirals and rapidly precessing spinning black hole binaries fall into that category. The inclination and polarization averaging we used for quasi-circular binaries is not applicable these systems, but  for simplicity we still  plot these signals against the standard sky and polarization averaged sensitivity curve. Writing the sky and polarization averaged signal-to-noise squared as
\begin{equation}
\overline{\rho}_{\Omega,\psi}^2 = 4 \int 
\frac{\vert \tilde{h}(f) \vert^2}{S_n(f)} \, df 
= \int_{f=0}^\infty 
 \frac{4 f^2  \vert  \tilde{h}(f) \vert^2}{(f S_n(f))} \, d(\ln f) \, ,
\end{equation}
indicates that an appropriate quantity to plot against the characteristic sensitivity  $(f S_n(f))^{1/2}$ is the dimensionless characteristic strain
\begin{equation}
h_c(f) = 2 f \left( \vert \tilde{h}_+(f) \vert^2+ \vert \tilde{h}_\times(f) \vert^2 \right)^{1/2}\, .
\end{equation}
We generate the Barycenter signals $h_+(t)$ and $h_\times(t)$, Fourier transform, and form $h_c(f)$.
To beautify the plots we smooth the numerically generated $h_c(f)$ using a running average over $\sim 100$ frequency bins. As an example, we generated augmented analytic kludge (AAK) EMRI waveforms~\cite{Chua:2017ujo} using the code provided at GitHub~\cite{emrigit} for a $20\, M_\odot$ stellar remnant black hole falling into a $10^6 M_\odot, \chi = 0.5$ spinning supermassive black hole at distance of 4 Gpc, starting 4 years before merger with an eccentricity of 0.5. This system has ${\rm SNR} = 52$, and the effective strain shown in Figure~\ref{fig:emri1}.

\subsection{Sky Dependent Estimates}

The sky averaged signal-to-noise ratios are useful for a first brush look at what systems might be detectable, but the signal-to-noise ratio can vary significantly across the sky, especially for short duration signals~\cite{Vallisneri:2012np}.
To incorporate the sky location dependence we must revert to using signals which have not been averaged over the sky location. The sky-location dependent SNRs are particularly useful for sources with known locations, such as the galactic verification binaries. We will continue to average over inclination angle and polarization angle as these are usually not well constrained.

The Michelson-type signal with spacecraft 1 at the vertex is given by
\begin{equation}
\fl \quad \quad s_{1}(t) = \frac{\delta \ell_{12}(t-2 L/c) + \delta \ell_{21}(t-L/c)}{2 L} - \frac{\delta \ell_{13}(t-2 L/c) + \delta \ell_{31}(t-L/c)}{2 L} \,\,,
\end{equation}
where the GW induced variation in LISA arm lengths between spacecraft $i$ and $j$ are given by
\begin{eqnarray}
 \frac{\delta \ell_{ij}(t)}{L} =& \frac{1}{2}(1+\cos^{2}\iota)d^{+}_{ij}(t) h^{+}(\xi_{i}) + \cos \iota \, d^{\times}_{ij}(t)h^{\times}(\xi_{i})   \,\,.
\end{eqnarray}
The detector terms $d^{+,\times}_{ij} = d^{+,\times}_{ij}\left(t; \textbf{x}_{i}, \textbf{x}_{j}, \theta, \phi, \iota, \psi\right)$~\cite{PhysRevD.69.082003} describes LISA's geometry through their dependence on the spacecraft position $\textbf{x}_{i}$. The variable $\xi_{i}$ define surfaces of constant gravitational wave phase at spacecraft $i$. The frequency domain representation of this signal can be found using the stationary phase approximation~\cite{PhysRevD.57.7089}

\begin{eqnarray}
\frac{\widetilde{\delta \ell}_{ij}(f)}{L} =& \left[\frac{1}{2}(1+\cos^{2}\iota)d^{+}_{ij}\left(t_{*}\right) + i\cos \iota \, d^{\times}_{ij}\left(t_{*}\right) \right] A(f) e^{i\left(\Psi(f) + \delta \Psi_{i}(f) \right)}
\end{eqnarray}
where $A(f)$ and $\Psi(f)$ are the amplitude and phase. The stationary time is given by the relation $t_{*}(f) = \Psi'(f)/2\pi$. This is used to map $d^{+}_{ij}(t)$ to $d^{+}_{ij}(f)$ {\it etc}. The motion of the LISA detector also imparts a phase shift $\delta \Psi_{i}(f) = 2\pi f \hat{\textbf{k}}\cdot\textbf{x}_{i}(t_*(f))/c$, where $\hat{\textbf{k}}$ defines the line of sight vector to the source. The PhenomA amplitude is given in equation (\ref{PhA}). The phase is written in terms of a power series expansion that is motivated the post Newtonian expansion, with additional terms that are found by fitting to numerical relativity simulations:
\begin{equation}
\fl \quad\quad  \Psi(f) = 2 \pi f t_0 +\phi_0 + \psi_0 f^{-5/3}+\psi_2 f^{-1} + \psi_3 f^{-2/3} + \psi_4 f^{-1/3}+\psi_6 f^{1/3}.
\end{equation}
The expansion coefficients $\psi_i$ depend on the masses and are given in terms of a numerical look-up table~\cite{Ajith:2007qp}. 

\begin{figure}[th]
	\centering
	\includegraphics[scale=1.0]{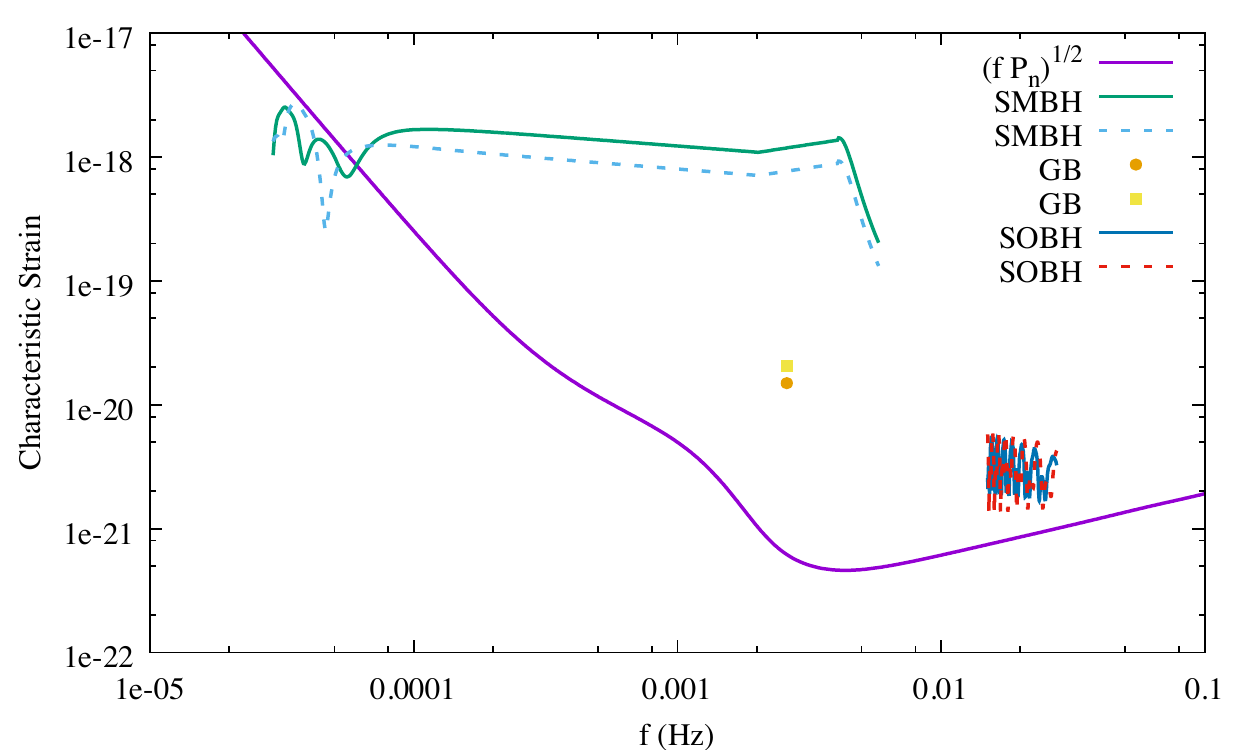} 
	\caption{\label{fig:SenseskyDep} The effective amplitudes of a SMBH, GB and SOBH are shown for two sky locations. Sine there is no sky averaging they are plotted against the characteristic noise $\sqrt{f P_{n}(f)}$. The masses are the same as for the sources shown shown in Figure~\ref{fig:all3} for the sky averaged case.}
\end{figure}

The average over the source inclination and polarization angles, $\left< |\tilde{s}_{1}(f)|^{2}\right>_{\iota, \psi}$, where the subscripts to the angle brackets denote which variables are averaged over, can be shown to be equivalent to computing the signal at two fixed values of inclination and polarization:
\begin{equation}
\left< |\tilde{s}_{1}(f)|^{2}\right>_{\iota, \psi} = \frac{8}{5}\left(  |\tilde{s}_{1}(f)|^{2}|_{\iota = \frac{\pi}{2}, \psi = 0}+ |\tilde{s}_{1}(f)|^{2}|_{\iota = \frac{\pi}{2}, \psi = \frac{\pi}{4}} \right) \, .
\end{equation}
This allows us to compute the orientation averaged signal using just two calls to the waveform generator. The sky location dependent signal-to-noise ratio is then
\begin{equation}\label{SNRtime}
\bar{\rho}_{\iota, \psi}^{2}(\theta,\phi) = 4 \int_{f=0}^{\infty} \frac{f^{2} \left< |\tilde{s}_{1}(f,\theta,\phi)|^{2}\right>_{\iota, \psi}}{f P_{n}(f)} d(\log f) \,\,.
\end{equation}
To give a visual impression of the signal strength we plot the characteristic strain $\tilde{h}_{\mathrm{eff}}(f) = 2 f \left< |\tilde{s}_{1}(f)|\right>_{\iota, \psi}$ against the characteristic noise amplitude in the Michelson channel $\sqrt{f P_{n}(f)}$. For a galactic binary we take a similar approach to calculate SNRs as for the sky averaged case:
\begin{equation}
\bar{\rho}_{\iota, \psi}^{2}(\theta,\phi) = 4 \int_{f_{\mathrm{in}}}^{f_{\mathrm{in}}+\Delta f} \frac{\left< |\tilde{s}_{1}(f)|^{2}\right>_{\iota, \psi}}{P_{n}(f)} d f \approx 4 \frac{\left< |\tilde{s}_{1}(f_{\mathrm{in}})|^{2}\right>_{\iota, \psi}}{P_{n}(f_{\mathrm{in}})} \Delta f  \,\,.
\end{equation}
An example of a sky-location dependent sensitivity plot is shown in Figure~\ref{fig:SenseskyDep} for the same sources shown previously in Figure~\ref{fig:all3}, but now at two different  sky locations ($\theta = 0.5,\phi = 2.3)$ and  ($\theta = 1.1,\phi = 1.5)$. The signal-to-noise ratio for these sky-dependent sources are 3106 and 2017 for the super massive black hole binary, 207 and 151 for the galactic binary, and 4.39 and 4.51 for the stellar origin black hole binary.

The oscillations in the tracks seen in Figure~\ref{fig:SenseskyDep} are due to the time dependent antenna pattern. For the supermassive black hole system most of the modulation is seen at low frequencies where the system spends many months. The evolution of the signal becomes far more rapid as it sweeps to higher frequencies, and the detector is effectively stationary on this timescales, so the amplitude no longer oscillates. The oscillations are very pronounced for the stellar origin black hole binary, which slowly evolves over the entire 4 year observation period.

\begin{figure}[th]
\centering	\includegraphics[scale=0.35]{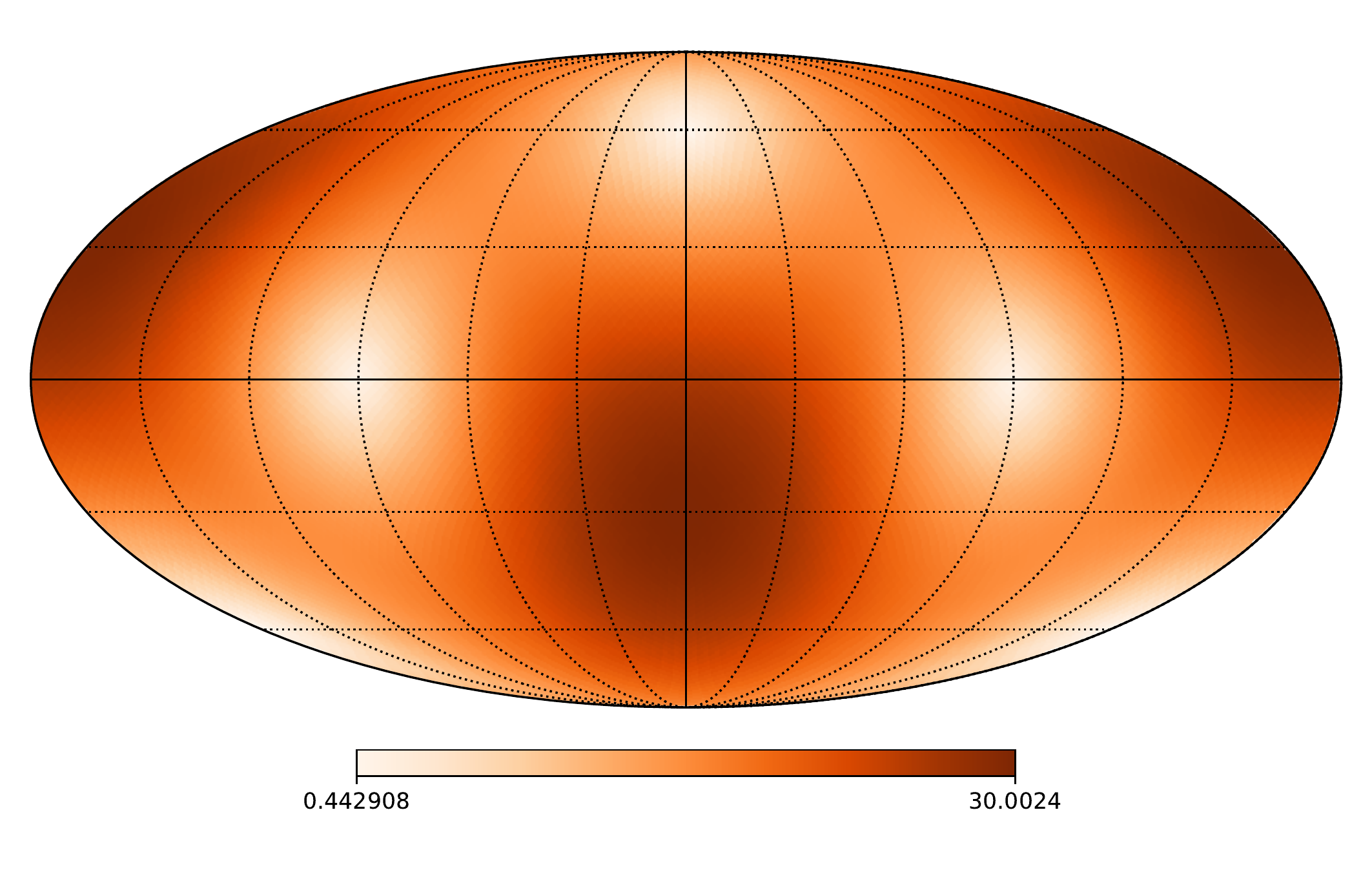} \includegraphics[scale=0.35]{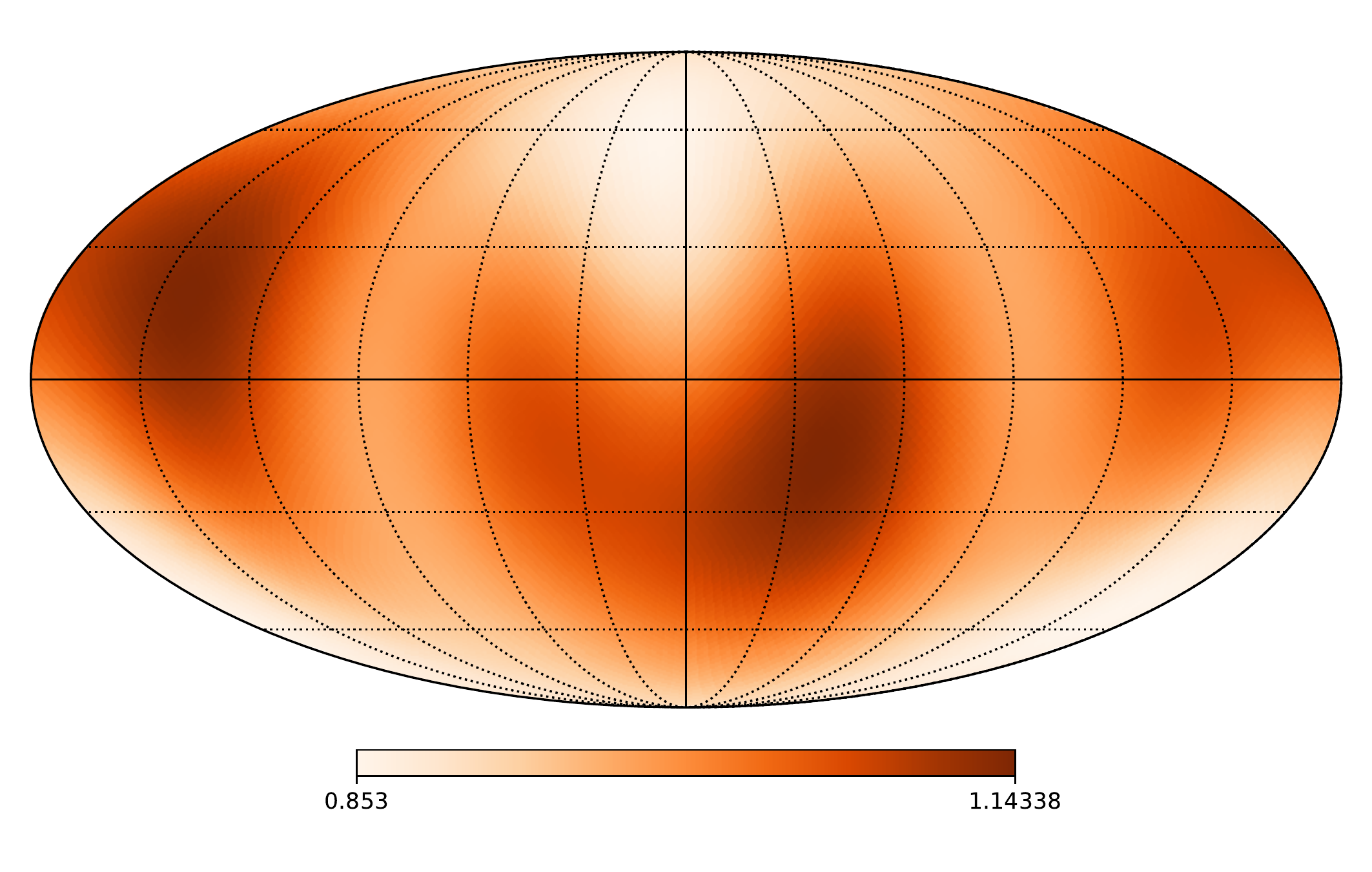} 
	\caption{\label{fig:skyMap} The sky location can greatly affect the signal-to-noise for a source. The maps show the SNR scaled by the all-sky average. The sky map on the left is for a $10^6 M_\odot$  black hole binary at $z=3$, while the map on the right is for a $10^4 M_\odot$ black hole binary at $z=0.6$.  Note the large difference in dynamic range for the two color maps. The standard deviation of the scaled SNR is 76\% for the $10^6 M_\odot$ system and 7\% for the $10^4 M_\odot$ mass system. The variance is smaller for lower mass systems as the signal accumulates more slowly over time and partially averages out the time variation of the LISA antenna pattern.}
\end{figure}

The sky location of the source plays a very large role in its detectability and our ability to characterize the source parameters. For example, the standard deviation of the SNR across the sky for the equal mass binary black hole system shown in Figure~\ref{fig:all3} is 76\% of the sky averaged value.The galactic binary and LIGO binary had deviations of 77\% and  39\% of the sky averaged value respectively. In Figure~\ref{fig:skyMap} we show the variation in the SNR across the sky for two binary black hole, one with detector frame total mass $4 \times 10^6 M_\odot$, and another with detector frame total mass $1.6 \times 10^4 M_\odot$. For the more massive system, most of the SNR is accumulated in the two days around merger, and since the detector is effectively stationary during this time, we recover the fixed quadrupolar antenna pattern. For the less massive system, where the merger lasts for a longer time, the sky map is more uniform.

\begin{figure}[th]
	\centering
	\includegraphics[scale=0.5]{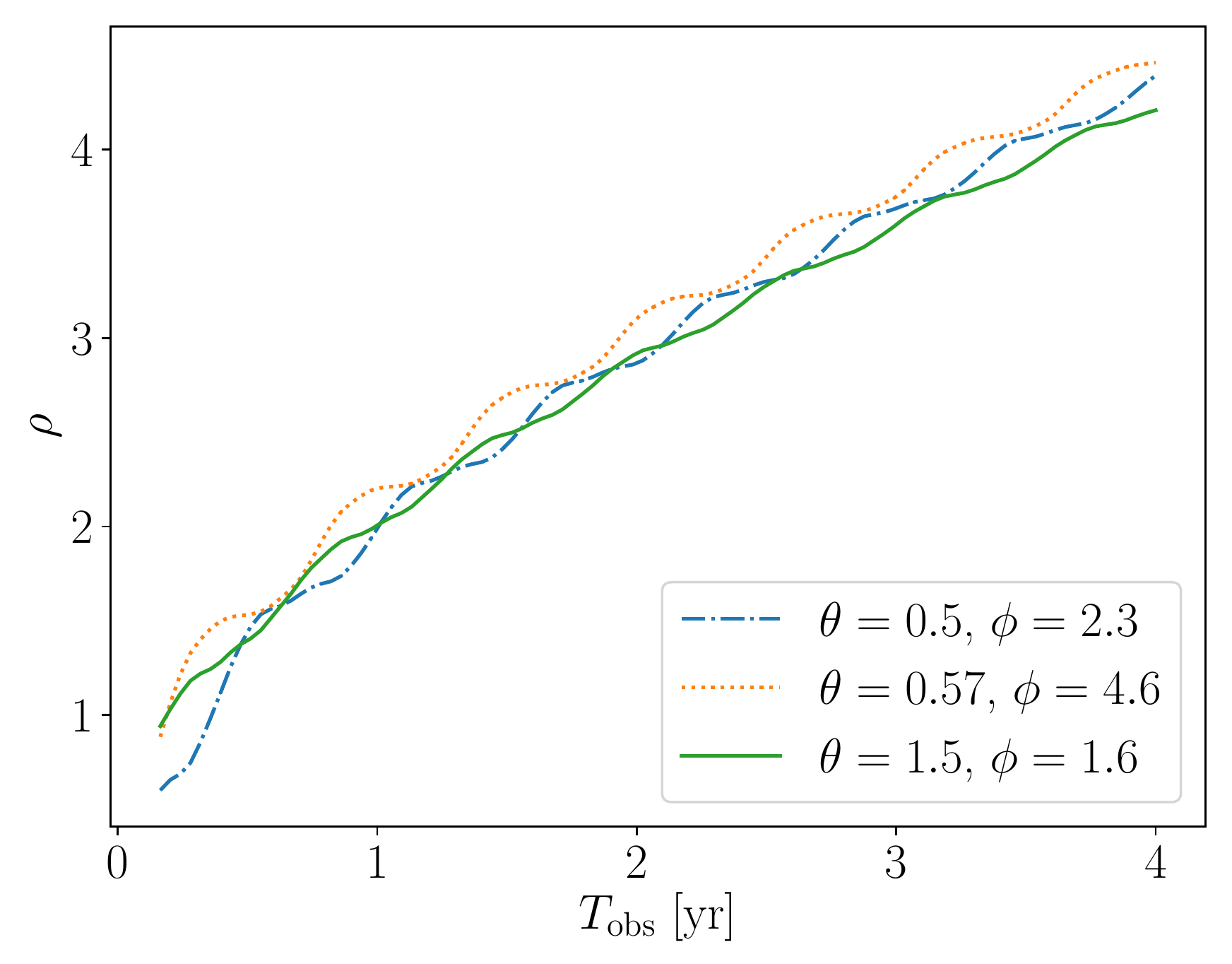} 
	\caption{\label{fig:SNR_time} For the LIGO binary this figure displays how the signal-to-noise grows with time depending on the sky location of the binary. The rate of SNR gain also depend on the sky location. The green line demonstrates a more uniform growth in SNR over time while the dotted orange line shows periods of quick growth followed by periods of slow growth.}
\end{figure}

Figure~\ref{fig:SNR_time} shows the signal-to-noise grows in time for the stellar origin black hole binary for various sky locations. We see that the rate of SNR growth depends on the sky location at any given time due to the changing orientation of the LISA antenna pattern. If the sources lies in a sensitive region of the antenna pattern then the SNR will grow quickly compared to when the source lies in an insensitive region.

An additional factor that will impact the SNR as a function of sky location is the time variation of the galactic confusion noise. The confusion noise is loudest when the most the peak of the antenna pattern sweeps across the galaxy, and quietest when pointed away from the galaxy. The confusion noise will vary adiabatically such that $S_c(f, t)$. Using the stationary phase approximation, the time dependence gets mapped to frequency dependence via $t_*(f)$, which allows for the time variation of the confusion noise to be incorporated in the SNR integral (\ref{SNRtime}). We defer an analysis of this effect to a future study.

\section{Acknowledgments}
We appreciate the input and feedback from Martin Hewitson, Emanuele Berti, Paul McNamara, and Davide Gerosa. TR and NJC appreciate the support of the NASA grant NNX16AB98G. CL acknowledges the support from the UCAS Joint PhD Training Program.

\bibliography{LISA_sense}

\end{document}